\documentclass[10pt]{iopart}
\begin{document}
\jl{3}
\title{Exciton trapping in magnetic wire structures}
\author{J A K Freire\dag \footnote[3]{Permanent address: Departamento de F\'{\i}sica,
Universidade Federal do Cear\'a, Centro de Ci\^encias, Caixa
Postal 6030, Campus do Pici, 60455-760 Fortaleza, Cear\'a,
Brazil. Electronic address: alking@fisica.ufc.br}, F M
Peeters\dag \footnote[4]{Electronic address:
peeters@uia.ua.ac.be}, V N Freire\ddag~ and G A Farias\ddag}
\address{\dag\ Departement Natuurkunde, Universiteit Antwerpen (UIA),
Universiteitsplein 1, B-2610 Antwerpen, Belgium}
\address{\ddag\ Departamento de F\'{\i}sica, Universidade Federal do Cear\'a, Centro de Ci\^encias,
Caixa Postal 6030, Campus do Pici, 60455-760 Fortaleza, Cear\'a,
Brazil}
\submitted
%
%
%
%
\begin{abstract}
The lateral magnetic confinement of quasi two-dimensional excitons
into wire like structures is studied. Spin effects are take into
account and two different magnetic field profiles are considered,
which experimentally can be created by the deposition of a
ferromagnetic stripe on a semiconductor quantum well with
magnetization parallel or perpendicular to the grown direction of
the well. We find that it is possible to confine excitons into
one-dimensional (1D) traps. We show that the dependence of the
confinement energy on the exciton wave vector, which is related
to its free direction of motion along the wire direction, is very
small. Through the application of a background magnetic field it
is possible to move the position of the trapping region towards
the edge of the ferromagnetic stripe or even underneath the
stripe. The exact position of this 1D exciton channel depends on
the strength of the background magnetic field and on the magnetic
polarisation direction of the ferromagnetic film.
\end{abstract}
\section{Introduction}
Lateral localization of excitons in semiconductors quantum wells
has been the subject of theoretical and experimental research in
the last few years. Such one dimensional traps can be realized,
e.g.,  by using nonhomogeneous stress to induce a hydrostatic
expansion which creates an energy minimum for the excitons
\cite{Nego}. Another way is to apply 1D spatial varying electrical
potentials in the plane of the quantum well and to use the
quantum-confined Stark effect in order to create a minimum in the
exciton effective potential \cite{Govorov,Zimmermann,Coco}.
Recent theoretical \cite{JAKF,JAKF2} studies on exciton
confinement have shown the feasibility of exciton trapping by
using circular (two dimensional) magnetic traps.

Due to recent advances in the design and manufacture of
microstructured magnetic potentials \cite{PeetersRev}, several
works have been published on the study of magnetic barriers and
magnetic superlattices. However, the majority of these
experimental and theoretical investigations study the influence of
such magnetic potentials on the properties of charged particles,
e.g., electrons \cite{Matulis,Ibrahim,Nogaret,BJ}.

In this work, we propose a natural extension of such magnetic
systems: the confinement of excitons in linear magnetic traps or
magnetic wire structures. This paper is structured as follows.
Sec. 2 gives a brief description of the magnetic wire structures.
The theoretical model for the description of the exciton motion
in a quantum well in the presence of linear magnetic traps is
given in Sec. 3. The discussion of the numerical results follows
in Sec. 4. As an example we show the results for
GaAs/Al$_{x}$Ga$_{1-x}$As quantum wells. Finally, a summary of
our results is presented in Sec. 5.
\section{Modelling the Magnetic Wire Structures}
Magnetic confinement potentials can be created by depositing
ferromagnetic stripes on top of a semiconductor quantum well. An
additional external homogeneous magnetic field can be applied to
the system in order to maximize the confinement. A sketch of the
experimental setup is shown in figure 1(a), and a schematic cross
section of the stripe with perpendicular and in-plane
magnetization can be seen in the left and right side of figure
1(b), respectively. In this system, only the $z$ component of the
magnetic field is responsible for the magnetic modulation, and can
be calculated through Coulomb's law \cite{Ibrahim}.

The corresponding equation for the $z$ component of the magnetic
field, assuming a stripe with magnetization perpendicular to the
$xy$ plane, located at $x ~\epsilon~ (-a/2,a/2)$, $y ~\epsilon~
(-\infty,\infty)$, and $z ~\epsilon~ (-h/2,h/2)$, where $a$ ($h$)
is the stripe width (thickness), can be written as follows:

\numparts
\begin{eqnarray}
B_z(x) &=& B_a + \frac{\mu_0 {\cal M}}{2\pi} \left[B_*(x,z+h/2) -
B_*(x,z-h/2)\right],
\end{eqnarray}
where
\begin{eqnarray}
B_*(x,z) &=& \left[\arctan\left(\frac{x+a/2}{z}\right) -
\arctan\left(\frac{x-a/2}{z} \right) \right].
\end{eqnarray}

For a stripe with in-plane magnetization the corresponding result
is
\begin{eqnarray}
B_z(x) &=& B_a + \frac{\mu_0 {\cal M}}{2\pi} \left[B_*(x+a/2,z) -
B_*(x-a/2,z)\right],
\end{eqnarray}
where
\begin{eqnarray}
B_*(x,z) &=& \frac{1}{2}\ln \left(
\frac{x^2+(z+h/2)^2}{x^2+(z-h/2)^2} \right).
\end{eqnarray}
\endnumparts

In the above equations, $z$ is the distance of the center of the
stripe to the middle of the quantum well, $\cal M$ is the stripe
magnetization, and $B_a$ the uniform applied magnetic field. The
corresponding magnetic field for the stripe with perpendicular
(solid curve) and in-plane (dashed curve) magnetization is shown
in figure 1(c). Notice that, in the case of a perpendicular
magnetization, the exciton will experience a positive effective
magnetic field underneath the stripe and negative elsewhere as
sketched on the left side of figure 1(b) [see also solid curve in
figure 1(c)]. The situation is completely different for the
stripe with in-plane magnetization, where now the effective
magnetic field has a negative peak in the region near the edge of
the stripe which corresponds to the direction of $\cal M$ (at $x
= a/2$) and a positive peak near the other (at $x = -a/2$) edge
[see right side of figure 1(b) and dashed curve in figure 1(c)].
\section{Theoretical Model}
The effective Hamiltonian describing the exciton motion in a
quantum well in a nonhomogeneous magnetic field, with isotropic
electron and heavy-hole effective masses $m_{e}^{\ast}$ and
$m_{h}^{\ast}$, respectively, can be written in the following way:
\numparts
\begin{equation}
H = H^{\bot }(z_{e},z_{h}) + W(r,z_{e},z_{h}) + H^{2D}({\mathbf
R},{\mathbf r}) + H^{m_z }(X), \label{Ham1}
\end{equation}
where
\begin{eqnarray}
H^{\bot }(z_{e},z_{h}) &=&-\frac{\hbar ^{2}}{2m_{e}^{\ast
}}\frac{\partial ^{2}}{\partial z_{e}^{2}}+V_{e}(z_{e}) -
\frac{\hbar ^{2}}{2m_{h}^{\ast }} \frac{\partial ^{2}}{\partial
z_{h}^{2}}+V_{h}(z_{h}),  \label{HZ1}
\end{eqnarray}
is the Hamiltonian describing the electron and heavy hole
confinement in the quantum well, i.e., in the $z$ direction;
\begin{eqnarray}
W(r,z_{e},z_{h}) &=&\frac{\gamma e^{2}}{\varepsilon
r}-\frac{e^{2}}{\varepsilon \sqrt{ r^{2}+(z_{e}-z_{h})^{2}}},
\label{HW1}
\end{eqnarray}
is related to the difference between the 2D and 3D Coulomb
interaction and is treated as a perturbation, where $\gamma$ is a
variational parameter which is chosen such that it minimizes
$W(r,z_{e},z_{h})$, as discussed in detail elsewhere \cite{Bao};
\begin{eqnarray}
\fl H^{2D}({\mathbf R},{\mathbf r}) &=& \frac{\hbar
^{2}}{2m_{e}^{\ast }} \left\{ -i \frac{m_{e}^{\ast }}{M}
\nabla_{R} - i \nabla_{r} + \frac{e}{\hbar c} A \left(X +
\frac{m_{h}^{\ast}}{M} x \right) \mathbf{e}_{y} \right\}^{2}
\nonumber \\ \fl &+& \frac{\hbar ^{2}}{2m_{h}^{\ast }} \left\{ -i
\frac{m_{h}^{\ast }}{M} \nabla_{R} - i \nabla_{r} - \frac{e}{\hbar
c} A \left(X - \frac{m_{e}^{\ast}}{M} x \right) \mathbf{e}_{y}
\right\}^{2} - \frac{\gamma e^{2}}{ \varepsilon r},  \label{H2D1}
\end{eqnarray}
describes the exciton motion in the $xy$ plane in a nonhomogeneous
magnetic field, which is described by the vector potential in the
Landau gauge ${\bf A}({\bf R})=A_{Y}(X){\bf e}_{Y}$. In writing
the above equation, we introduced the exciton relative and
center-of-mass motion coordinates, ${\mathbf r} = {\mathbf r}_{e}
- {\mathbf r}_{h}$ and $\ {\mathbf R} = \left(m_{e}^{\ast}{\mathbf
r}_{e} + m_{h}^{\ast}{\mathbf r}_{h}\right) / M $, respectively,
with the total exciton mass $ M = (m_{e}^{\ast}+m_{h}^{\ast})$.
Finally, the last term in equation (\ref{Ham1}), i.e.,
\begin{eqnarray}
H^{m_z }(X) &=& \mu _{B}\left[g_{e,z}S_{e,z}-\frac{1}{3}
g_{h,z}J_{h,z}\right] B_{z}(X) - \frac{2}{3}
\sum_{i=1}^{3}c_{i}S_{e,i}{}J_{h,i},  \label{HSpin1}
\end{eqnarray}
\endnumparts
is the Hamiltonian describing the exciton spin interaction with
the nonhomogeneous magnetic field, which is oriented along the $z$
direction. We consider here that the contribution of the spin
interaction with the in-plane magnetic field is very small and
can be neglected \cite{Kest,Snel94}. In writing down the above
spin Hamiltonian, the adiabatic approximation was used, which
assumes that the relative motion is fast as compared to the
center-of-mass motion. This allows us to expand the magnetic field
to zero order in the relative motion coordinate \cite{JAKF2}. In
Eq. (2e), $\mu _{B}=e\hbar /2m_{e,\Vert }^{\ast }c$ is the Bohr
magneton, $m_z = S_{e,i} + J_{h,i}$ is the total exciton spin
quantum number, which is related to the electron ($S_{e,i} = \pm
1/2$) and heavy-hole ($J_{h,i} = \pm 3/2$) spin quantum numbers,
$c_{i}$ is the spin-spin coupling constant related to the zero
field spin interaction, and $g_{e,z}$, $g_{h,z}$ are the $z$
component of the electron and heavy-hole $g$ factor, respectively
\cite{Snel94}.

Following the approach of Freire {\it et al} \cite{JAKF}, the
$H^{2D}({\mathbf R},{\mathbf r})$ Hamiltonian [see equation
(\ref{H2D1})] can be simplified by using a transformation of the
wave function analogous to the one used in the homogeneous
magnetic field case, which assumes the existence of an exact
integral of motion, namely the magnetic center-of-mass momentum
\cite{Gor}:
\begin{eqnarray}
\Psi ^{m_z }({\mathbf R},{\mathbf r},z_{e},z_{h}) &\Rightarrow&
\exp \left[ -\frac{ie}{\hbar c}yA_{Y}(X)\right] \Psi ^{m_z
}({\mathbf R},{\mathbf r},z_{e},z_{h}). \label{wavetrans}
\end{eqnarray}

Further, we can use the adiabatic approximation to expand the
vector potential up to the second order in the relative motion
coordinates, i.e.,
\begin{eqnarray}
A_{Y}\left(X\pm\frac{m_{h(e)}^*}{M}x\right) = A_{Y}\left( X\right)
\pm \frac{ m_{h(e)}^*}{M}xB_{z}(X). \label{vectexpans}
\end{eqnarray}

Inserting equations (\ref{wavetrans}) and (\ref{vectexpans}) into
equation (\ref{H2D1}), the $H^{2D}$ Hamiltonian can be separated
in relative and center-of-mass motion Hamiltonians, i.e.,
$H^{2D}({\mathbf R},{\mathbf r}) = H^{rel}({\mathbf r},{\mathbf
R},\nabla_{R}) + H^{CM}({\mathbf R})$, with $H^{rel}({\mathbf
r},{\mathbf R},\nabla_{R}) = - ({\hbar^{2}}/ {2\mu})
\nabla_{r}^{2} - ({\gamma e^{2}}/{\varepsilon r}) + u_1 + u_2$,
where
\begin{eqnarray}
\fl u_1 = - \frac{ie\hbar }{\mu c}\xi B_{z}(X)x\frac{\partial
}{\partial y} - \frac{ie\hbar }{Mc}\left( B_{z}(X)\frac{\partial
}{\partial Y}\right) x \nonumber +
 \frac{ie\hbar }{Mc}\left(\frac{\partial
}{\partial X} B_{z}(X)+B_{z}(X)\frac{\partial }{\partial X}\right)
y, \\ \fl u_2 = \frac{e^{2}}{2Mc^{2}}B_{z}(X)^{2} \left(\frac{\xi
^{2}M}{\mu}x^{2} + r^2\right), \label{W1W2}
\end{eqnarray}
and $H^{CM}({\mathbf R}) = -({\hbar ^{2}}/{2M})\nabla_{R}^{2}$.
Here, $\mu = m_{e}^{*} m_{h}^{*} / M $ is the exciton reduced
mass, and $ \xi = (m_{h}^{*} - m_{e}^{*}) / M $. The procedure
used to go from equation (\ref{H2D1}) to equation (\ref{W1W2}) is
analogous to the one used in our previous work [see equations
(1)-(7) in reference \cite{JAKF}] on the trapping of excitons in
circular magnetic traps.

To find the eigenvalues of the Hamiltonian, equation (\ref{Ham1}),
we use the adiabatic approximation to decouple the exciton
center-of-mass motion, which is slow, from the exciton relative
motion, which is fast. We already assumed that the spin degrees of
freedom depend only on the center-of-mass coordinates. Thus, the
total exciton wave function becomes a product of the wave
function of the decoupled motions:
\begin{eqnarray}
\Psi ^{m_z }({\mathbf R},{\mathbf r},z_{e},z_{h})= \varphi
({\mathbf R}) \Phi({\mathbf r})F (z_{e},z_{h}){\cal L}^{m_z
}({\mathbf R}).
\end{eqnarray}

The energy of the exciton confinement in the quantum well ($z$
direction), i.e., the  eigenvalue of $\{ H^{\bot }(z_{e},z_{h}) -
E^{\bot }\}F (z_{e},z_{h}) = 0$, does not dependent on the
magnetic field and therefore will not contribute to the magnetic
exciton trapping energy. Therefore, this energy does not have to
be calculated explicitly. In order to solve the variational
equation $\partial \triangle E^{'}/\partial \gamma  = 0$,  with $
\triangle E^{'} = \left\langle \Phi ({\mathbf r}) F
(z_{e},z_{h})\left| W(r,z_{e},z_{h})\right| \Phi ({\mathbf r}) F
(z_{e},z_{h})\right\rangle $, to calculate $\gamma$ we follow the
same procedure as described in our previous work \cite{JAKF2}.

The eigenvalues and eigenfunctions of the equation for the exciton
relative motion $\{ H^{rel}({\mathbf r},{\mathbf R},\nabla_{R}) -
E^{rel}\}\Phi({\mathbf r}) = 0$ are calculated using a
perturbation technique, i.e., all the $B$-dependent terms [see
$u_1$ and $u_2$ in equation (\ref{W1W2})] are treated as
perturbations. A full discussion of the calculation method for
the relative motion energy is given in reference \cite{JAKF} and
will not be repeated here. The spin energies can be easily
calculated. The eigenvalues of $\{H^{m_z }(X) - E^{m_z}\} {\cal
L}^{m_z }({\mathbf R}) = 0$ can be written as $E^{m_z} = \pm
\frac{1}{2}\mu_{B} \left[(-1)^{m_z+1}g_{e,z} + g_{h,z}\right]
B_{z}(X)$ \cite{JAKF2}. In writing  $E^{m_z}$, we neglected the
terms which do not depend on the magnetic field ($\pm c_z/2$), and
the terms with $c_x$, and $c_y$ cancel because of symmetry
considerations. The different sign in the above spin energy are
related to the total spin quantum number $m_z = \pm 1$ (which is
connected with the $\sigma^{\pm}$ polarized states) and $ m_z =
\pm 2$ (which is related to the dark excitons) \cite{Kest}. In
this work, the $ m_z = \pm 2$ states will not be considered
because they are not optically active.

Finally, we obtain the following Schr\"{o}dinger-like equation
for the exciton center-of-mass motion:
\numparts
\begin{eqnarray}
\left\{ -\frac{\hbar ^{2}}{2}\frac{d}{dX}\left[
\frac{1}{M^{eff}(X)}\frac{d}{dX}\right] +V^{eff}(X)-E\right\} \psi
(X)=0.
\end{eqnarray}

The above equation includes the different eigenvalues for the
fast motion which contribute to an effective potential and also
results in a spatial dependent effective mass:
\begin{eqnarray}
\fl M^{eff}(X) / M &=&
\left[1-\frac{e^{2}\mu}{\hbar^{2}M^{2}c^{2}}
\frac{\alpha_{m_{r}}^{n_{r}}} {\gamma^4}B_{z}(X)^{2}\right]^{-1},
\label{effecmass} \\ \fl V^{eff}(X) &=& \frac{\hbar ^{2}}{2
M^{eff}(X)}Q_{Y}^{2} + \frac{e^{2}\zeta}{2\mu c^{2}}
\frac{\beta_{m_{r}}^{n_{r}}}{\gamma^2} B_{z}(X)^{2} \nonumber \\
&+& \frac{e\hbar} {2\mu c}\xi B_{z}(X)m_{r} \pm \frac{1}{2}\mu_{B}
\left[(-1)^{m_z+1}g_{e,z} + g_{h,z}\right] B_{z}(X) .
\label{effecpot}
\end{eqnarray}
\endnumparts

We decoupled the wave function of the center-of-mass motion
assuming that $\varphi ({\mathbf R}) = \exp(iQ_{Y}Y) \psi(X)$,
where $Q_Y$ is the wave vector of the center-of-mass motion in
the $Y$ direction (free direction in the wire). Here, $\alpha
_{m_{r}}^{n_{r}}$ and $ \beta _{m_{r}}^{n_{r}}$ (in units of
${a_{B}^{\ast}}^4$ and ${a_{B}^{\ast}}^2$, respectively) are
constants related to the relative radial ($n_{r}$) and angular
($m_{r}$) quantum numbers \cite{JAKF}, and $\zeta = ({m_e^*}^2 +
{m_h^*}^2) / M$. Notice that in contrast to the case of exciton
trapping through 1D scalar potentials created by an electric
field \cite{Govorov}, the effective potential for linear magnetic
traps depends on the exciton wave vector $Q_Y$ [see first term in
equation (\ref{effecpot})]. The main contribution for the exciton
confinement is given by the two last terms in the effective
potential equation (\ref{effecpot}), which are the orbital and
spin Zeeman terms, respectively. The latter is a sensitive
function of the exciton $g$ factor \cite{Snel94}. The diamagnetic
contribution for the exciton effective potential [see second term
in equation (\ref{effecpot})] contains the contribution of the
well confinement through the parameter $\gamma$, which can be very
important for exciton trapping when the orbital and the spin
contribution are negligible, i.e., when the exciton angular
momentum is zero and the exciton $g$ factor is small \cite{JAKF2}.
\section{Exciton Trapping}
The trapping energy is defined as the difference in exciton energy
for an exciton in a homogeneous applied field $B_a$ and the
corresponding state in the nonhomogeneous magnetic field. For our
numerical calculations we used the following parameters for the
ferromagnetic stripe: width $a = 0.50$ $\mu$m, thickness $h =
0.20$ $\mu$m, distance to the center of the quantum well $d =
0.08$ $\mu$m, and the value of the stripe magnetization
corresponding to iron ${\cal M} = 1740$ emu/cm$^{3}$
\cite{Nogaret,Kitel}. We used for the GaAs/Al$_{0.3}$Ga$_{0.7}$As
quantum well an effective mass $m_e^* = 0.067 m_0$ for electrons
and $m_h^*=0.34 m_0$ for heavy-holes ($m_0$ is the free space
electron mass), and the GaAs dieletric constant $\varepsilon =
12.53$. The different $g$ factors are taken from reference
\cite{Snel94} and depend on the width of the GaAs quantum well.

The effective potential and the corresponding effective mass for
the exciton confinement [see equation (\ref{effecpot}) and
(\ref{effecmass}), respectively] for the exciton 1s ground state
are shown in figure \ref{figure2}(a) as a function of the
center-of-mass $X$ coordinate for the stripe with perpendicular
magnetization, and in figure \ref{figure2}(b) for the stripe with
in-plane magnetization. In these figures, we took a quantum well
width $L = 80$ {\AA}, an applied magnetic field $B_a = 0.15$ T,
and spin quantum numbers $m_z = -1$ (dotted curve) and $m_z = +1$
(dashed curve) for an exciton wave vector $Q_Y = 0$ $\mu$m$^{-1}$,
and $m_z = -1$ (dashed-dotted-dotted curve), $m_z = +1$
(dashed-dotted curve) for $Q_Y = 5$ $\mu$m$^{-1}$. Notice that the
effective potentials for $Q_Y = 0$ and $Q_Y = 5$ $\mu$m$^{-1}$ are
only slightly different, which suggests that the corresponding
energies are close to each other and that the contribution of the
$Q_Y$-dependent term to the exciton confinement potential [see
first term in equation (\ref{effecpot})] is small. Also notice
that for the case of perpendicular magnetization, the effective
potential for $m_z \leq 0$ has two minima near the edge of the
magnetic stripe, which will be the position where the exciton will
be magnetically trapped. For the parallel magnetization case and
also for the perpendicular one when $m_z > 0$, the potential
minimum occurs underneath the magnetic stripe and the exciton
localizes exactly under the magnetic stripe. It is worthwhile to
point out that for a given magnetization, the exciton spin
interaction with the nonhomogeneous magnetic field can be
responsible for a displacement in the minimum of the effective
potential related to the different $\sigma^\pm$ polarized states,
which is due to a change in the sign of the spin contribution [see
last term in equation (\ref{effecpot})].

The trapping energy of the exciton ground state as a function of
the exciton wave vector $Q_Y$ is shown in figure \ref{figure3}
for an applied field of $B_a = 0.15$ T and a well width $L = 80$
{\AA}, for the stripe with perpendicular [figure \ref{figure3}(a)
for $m_z=+1$ and \ref{figure3}(b) for $m_z=-1$] and in-plane
[figure \ref{figure3}(c) for $m_z=+1$ and \ref{figure3}(d) for
$m_z=-1$] magnetization. The trapping energy has a parabolic
dispersion as expected for free particle like motion [see equation
(\ref{effecpot})]. The effect of the center-of-mass momentum $Q_Y$
on the exciton trapping energy is very small. Notice that the
term $(\hbar^2/2M)Q_{Y}^{2}$, which does not depend on the
magnetic field, is not included in the calculation of the
trapping energy. Only the contribution due to the nonhomogeneous
magnetic field is shown in figure \ref{figure3}.

Experimentally, it is very useful to control the exciton
localization by means of tunable parameters, e.g., electric and
magnetic fields, without the necessity of changing the structural
parameters (e.g., quantum well width). For that purpose, we
analyzed the dependence of the exciton trapping energy under the
influence of an external applied homogeneous magnetic field $B_a$,
which is able to change completely the effective potential
responsible for the trapping of the exciton. Figures
\ref{figure4} and \ref{figure5} show the exciton trapping energy
as a function of the applied field $B_a$, for the stripe with
perpendicular and in-plane magnetization, respectively, for the
exciton 1s ground state. Results are given for quantum well
widths $L$ 50 {\AA} (a), 100 {\AA} (b), and 150 {\AA} (c). The
exciton is always trapped in the case $m_z = +1$, except for a
quantum well width $L = 100$ {\AA}. The reason is that the
corresponding exciton $g$ factor for this well width is almost
zero \cite{Snel94}, which greatly decreases the importance of the
spin term in the effective potential expression [see equation
(\ref{effecpot})]. This is also the explanation for the large
difference in magnitude (2 orders in some cases) of the energy
scale for $L = 50$ and $150$ {\AA} as compared to the $L = 100$
{\AA} situation.

The trapping energy of the stripe with in-plane magnetization
(figure \ref{figure5}) has a completely different behavior from
the one with perpendicular magnetization. First of all, the
energy is an even function of the applied field $B_a$ due to the
functional behavior of the corresponding nonhomogeneous magnetic
field [see dashed line in figure \ref{figure1}]. For certain
quantum well widths there exists also a critical field $B_a$
below which the exciton is not trapped. Changing the spin quantum
number from $m_z = +1$ into $m_z = -1$ drastically changes the
exciton confinement dependence on the quantum well width for both
magnetization directions. Using these results, one can choose an
optimal situation in which the effect of the nonhomogeneous
magnetic field on the exciton confinement is maximal and can be
detected by photoluminescence (PL) experiments. By changing the
applied magnetic field strength, one can modify the position of
the line spectra in the PL experiments related to the $\sigma^\pm$
polarized states in such a way that only one of them exist (the
other will be a dark state), which may make the effect of the
magnetic field on the exciton confinement easier to detect.

Due to the competition between the diamagnetic interaction, the
spin interaction and the applied magnetic field, the exciton can
be confined in different regions of space. This is illustrated in
figures \ref{figure6} and \ref{figure7}, where we show the wave
function for the exciton ground state as a function of the $X$
coordinate, for the stripe with perpendicular and in-plane
magnetization, respectively, for spin quantum numbers (a)
$m_z=-1$ and (b) $m_z=+1$, and for different values of the
applied field $B_a$. Notice that there is no confined exciton for
the $m_z = -1$ state when $B_a = 0$, which is not the case for
the $m_z = +1$ state. This can be easily explained by looking at
the corresponding effective potentials [see dotted curves in the
inset of figures \ref{figure6} and \ref{figure7}]. There is no
confinement region for the exciton when $m_z = -1$. Notice that
for the $m_z = -1$ ($m_z = +1$) situation [see figure
\ref{figure6}(a) and \ref{figure6}(b), respectively], the wave
function is located near the two edges of the ferromagnetic
stripe for $B_a$ positive (negative) and underneath the stripe
for $B_a$ negative (positive). For the in-plane magnetization
situation [figure \ref{figure7}], the exciton is confined at one
of the edges of the stripe for $m_z = -1$. Changing the sign of
the background field switches the exciton from one side of the
ferromagnetic edge to the other. A similar behavior is found for
$m_z = +1$ but: (i) the position of the exciton for a given $B_a$
is at the opposite edge from the one for $m_z = -1$; and (ii) for
$B_a = 0$ T, the exciton is bound and its wave function is
symmetric. This magnetic field dependence is consistent with the
$B_a \rightarrow - B_a$ symmetry of the binding energy [see
figure \ref{figure5}].
\section{Conclusion}
We studied the behavior of quasi-2D excitons in the presence of 1D
magnetic traps. In this system, the excitons are confined into
quantum wire like states. Such linear magnetic potentials can be
created experimentally by the deposition of a ferromagnetic
stripe on top of a semiconductor quantum well with an additional
external magnetic field applied perpendicularly to the stripe in
order to maximize the exciton confinement. In this paper, we
studied both possibilities of the stripe magnetization, i.e,
magnetization parallel or perpendicular to the plane of the
quantum well.

The main conclusions can be summarized as follows: (i) excitons
can be trapped by linear magnetic traps; (ii) the dependence of
the confinement energy on the exciton wave vector is small, but
the exciton confinement is very sensitive to its spin orientation;
and (iii) the confinement of excitons in such systems can be
controlled by an external tunable parameter, i.e., an applied
magnetic field. By changing the strength of the magnetic field one
can move the confinement region from underneath the magnetic
stripe to its edges or even to one of the edges. For the case of
parallel magnetization, the external magnetic field can move the
confinement region from one edge of the magnetic stripe to the
other edge of the stripe. The numerical obtained trapping energies
are rather small but they can be increased substantially if the
amplitude of the modulated magnetic field (see figure 1(c)) is
enhanced.

We also showed that the applied magnetic field can make a bright
exciton becomes a dark one. This transformation is very sensitive
to the quantum well width, the magnetization of the magnetic
stripe, and the exciton spin contribution, i.e. through the value
of the effective g-factor. Our results can help the design of
experiments to probe the exciton trapping by giving a very good
estimate of the set of parameters for which the confined exciton
can be detected, i.e., where the exciton is actually trapped in
the magnetic wire-like confinement potential. Further, the spatial
displacement of the exciton localization as controlled by changing
the external applied magnetic field, from e.g. one edge of the
magnetic stripe to the other, can be used experimentally to detect
and analyze the effect of the nonhomogeneous magnetic field on the
exciton motion. Our work is expected to provide useful insights
and stimulate further developments in experimental work involving
trapping of excitons in nonhomogeneous magnetic fields.
\ack%
{This research was partially supported by FWO-Vl, IUAP (Belgium),
the "Onderzoeksraad van de Universiteit Antwerpen (UIA)", and by
the Inter-university Micro-Electronics Center (IMEC, Leuven). J.
A. K. Freire was supported by the Brazilian Ministry of Culture
and Education (MEC-CAPES) and F. M. Peeters is a research director
with the FWO-Vl. V. N. Freire and G. A. Farias received financial
support from the Brazilian National Research Council (CNPq) and
the Brazilian Ministry of Planning through FINEP.}
\Figures
\begin{figure}
\caption{(a) Sketch of the experimental configuration showing the
structural parameters: the stripe width $a$, the thickness $h$,
and the distance of the stripe to the middle of the quantum well
$d$. The applied magnetic field is $B_a$, the stripe magnetization
$\cal{M}$, and the well width $L$. (b) Schematic cross section of
the stripe together with the magnetic field lines for
magnetization perpendicular and parallel to the $x$ direction. (c)
Magnetic field profile corresponding to the stripe with
perpendicular (\full) and parallel (\dashed) magnetization. The
shaded area is the position of the magnetic stripe.}
\label{figure1}
\end{figure}
\begin{figure}
\caption{Effective potential for the exciton 1s ground state as a
function of the center of mass of the exciton $X$, in case of a
magnetic stripe with magnetization (a) perpendicular and (b)
parallel to the $X$ direction. In this figure, the quantum well
width is $L = 80$ {\AA}, the applied magnetic field $B_a = 0.15$
T, the total spin quantum numbers $m_z = -1$ (\dotted) and $m_z =
+1$ (\dashed) for $Q_Y = 0$ $\mu$m$^{-1}$, and $m_z = -1$
(\dashddot), $m_z = +1$ (\chain) for $Q_Y = 5$ $\mu$m$^{-1}$. The
corresponding effective mass is also shown (\full).}
\label{figure2}
\end{figure}
\begin{figure}
\caption{Contribution of the ground state exciton trapping energy
to the wave vector $Q_Y$ dependence for: the stripe with
magnetization perpendicular to the $X$ axis, and spin numbers (a)
$m_z = +1$ and (b) $m_z = -1$; the stripe with parallel
magnetization, and spin numbers (c) $m_z = +1$ and (d) $m_z =
-1$. The quantum well width $L = 80$ {\AA}, and the applied
magnetic field $B_a = 0.15$ T.} \label{figure3}
\end{figure}
\begin{figure}
\caption{Trapping energy of the exciton ground state as a
function of the applied magnetic field $B_a$ for the stripe with
magnetization perpendicular to the $X$ direction and  well widths
(a) $L = 50$ {\AA}, (b) $L = 100$ {\AA}, (c) $L = 150$ {\AA}. The
spin numbers $m_z = -1$ (\dashed) and $m_z =+1$ (\full), and $Q_Y
= 0$ $\mu$m$^{-1}$.} \label{figure4}
\end{figure}
\begin{figure}
\caption{The same as figure \ref{figure4} but the stripe has
magnetization parallel to the $X$ axis.} \label{figure5}
\end{figure}
\begin{figure}
\caption{Wave function of the exciton ground state as a function
of $X$,  for the stripe with magnetization perpendicular to the
$X$ direction, with spin quantum numbers (a) $m_z = -1$ and (b)
$m_z = +1$. The quantum well width $L = 80$ {\AA}, the wave
vector $Q_Y = 0$ $\mu$m$^{-1}$, and the applied magnetic field
$B_a = -0.15$ T (\dashed), $B_a = 0$ T (\dotted), and $B_a =
0.15$ T (\full). The inset shows the respective effective
potentials.} \label{figure6}
\end{figure}
\begin{figure}
\caption{{The same as figure \ref{figure6}, but the stripe has
magnetization parallel to the $X$ axis.} \label{figure7}}
\end{figure}

\section*{References}
\begin{thebibliography}{9}
%
\bibitem{Nego} Negoita V, Snoke D W and Eberl K 1999 {\it Appl. Phys. Lett.} {\bf 75}
2059
%
\bibitem{Govorov} Govorov A O and Hansen W 1998 \PR B {\bf58} 12980
%
\bibitem{Zimmermann} Zimmermann S, Schedelbeck G, Govorov A O,
Wixforth A, Kotthaus J P, Bichler M, Wegscheider W and Abstreiter
G 1998 {\it Appl. Phys. Lett.} {\bf 73} 154
%
\bibitem{Coco} Cocoletzi G H and Ulloa S E 2000 \PR B {\bf 61}
13099
%
\bibitem{JAKF} Freire J A K, Matulis A, Peeters F M, Freire V N and Farias G
A 2000 \PR B {\bf 61} 2895
%
\bibitem{JAKF2} Freire J A K, Matulis A, Peeters F M, Freire V N and Farias G
A 2000 \PR B {\bf 62} 7316
%
\bibitem{PeetersRev} For a recent review see: Peeters F M and De
Boeck J 1999 {\it Handbook of nanostructured materials and
nanotechnology} vol 3 ed H S Nalwa (New York: Academic Press)
p~345
%
\bibitem{Matulis} Peeters F M and Matulis A 1993 \PR B {\bf 48} 15166
%
\bibitem{Ibrahim} Ibrahim I S and Peeters F M 1995 \PR B {\bf 52} 17321
%
\bibitem{Nogaret} Nogaret A, Bending S J and Henini M 2000 \PRL {\bf 84} 2231
%
\bibitem{BJ} Reijniers J and Peeters F M 2000 \JAP {\bf87} 8088
%
\bibitem{Bao} Wei B -H, Liu Y, Gu S -W and Yu K -W 1992 \PR B {\bf 46} 4269
%
\bibitem{Kest} van Kesteren H W, Cosman E C, van der Poel W A J A
and Foxon C T 1990 \PR B {\bf 41} 5283
%
\bibitem{Snel94} Blackwood E, Snelling M J, Harley R T, Andrews S R and Foxon C T
B 1994 \PR B {\bf 50} 14246
%
\bibitem{Gor} Gor'kov L P and Dzyaloshinsky I E 1968 {\it Sov. Phys. JETP} {\bf 26} 449
%
\bibitem{Kitel} Kittel C 1976 {\it Introduction to Solid State
Physics} (New York: J Wiley)
%
\end{thebibliography}
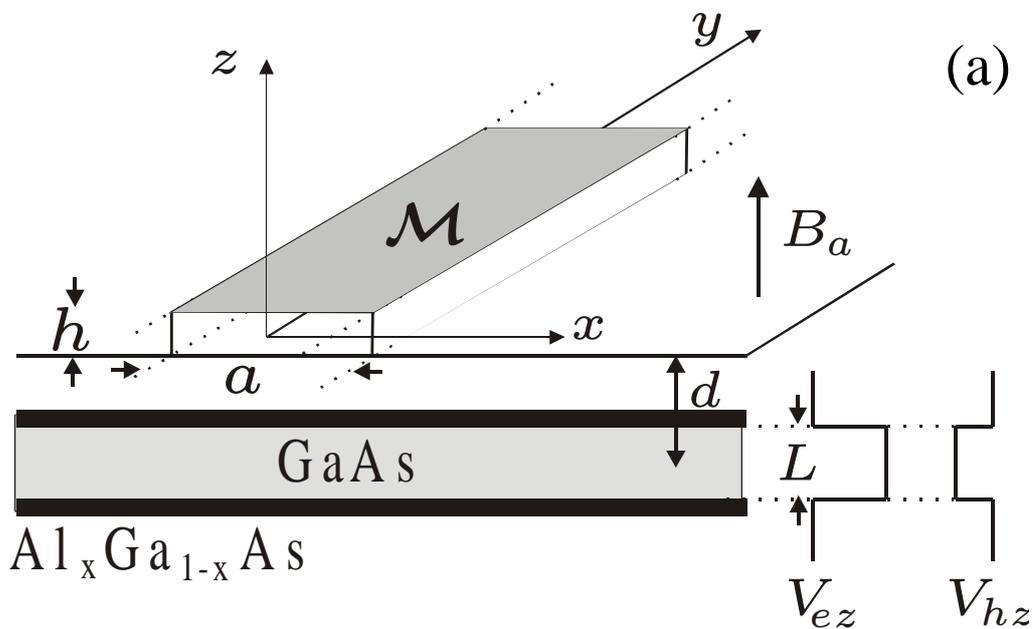
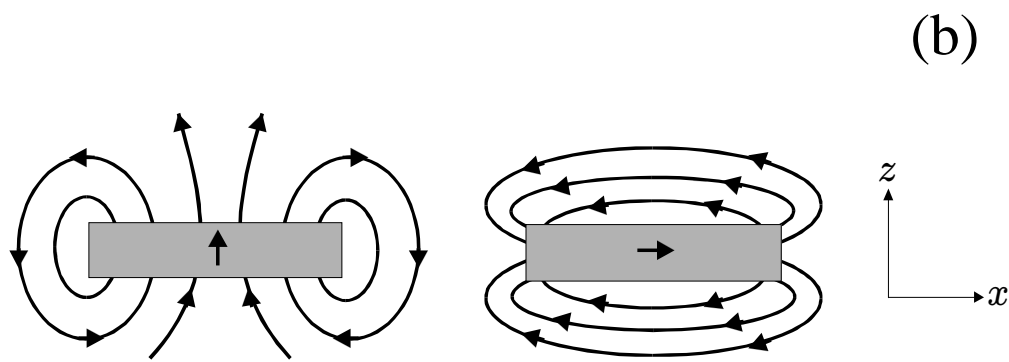
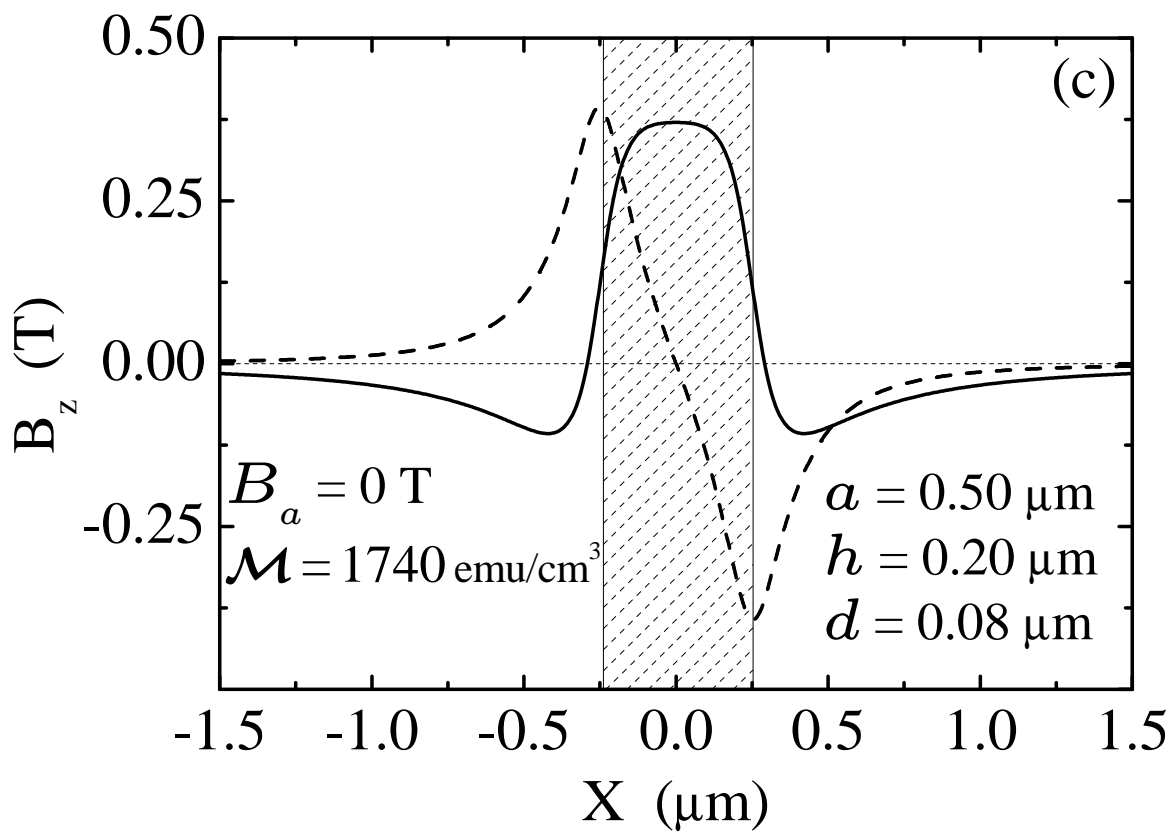
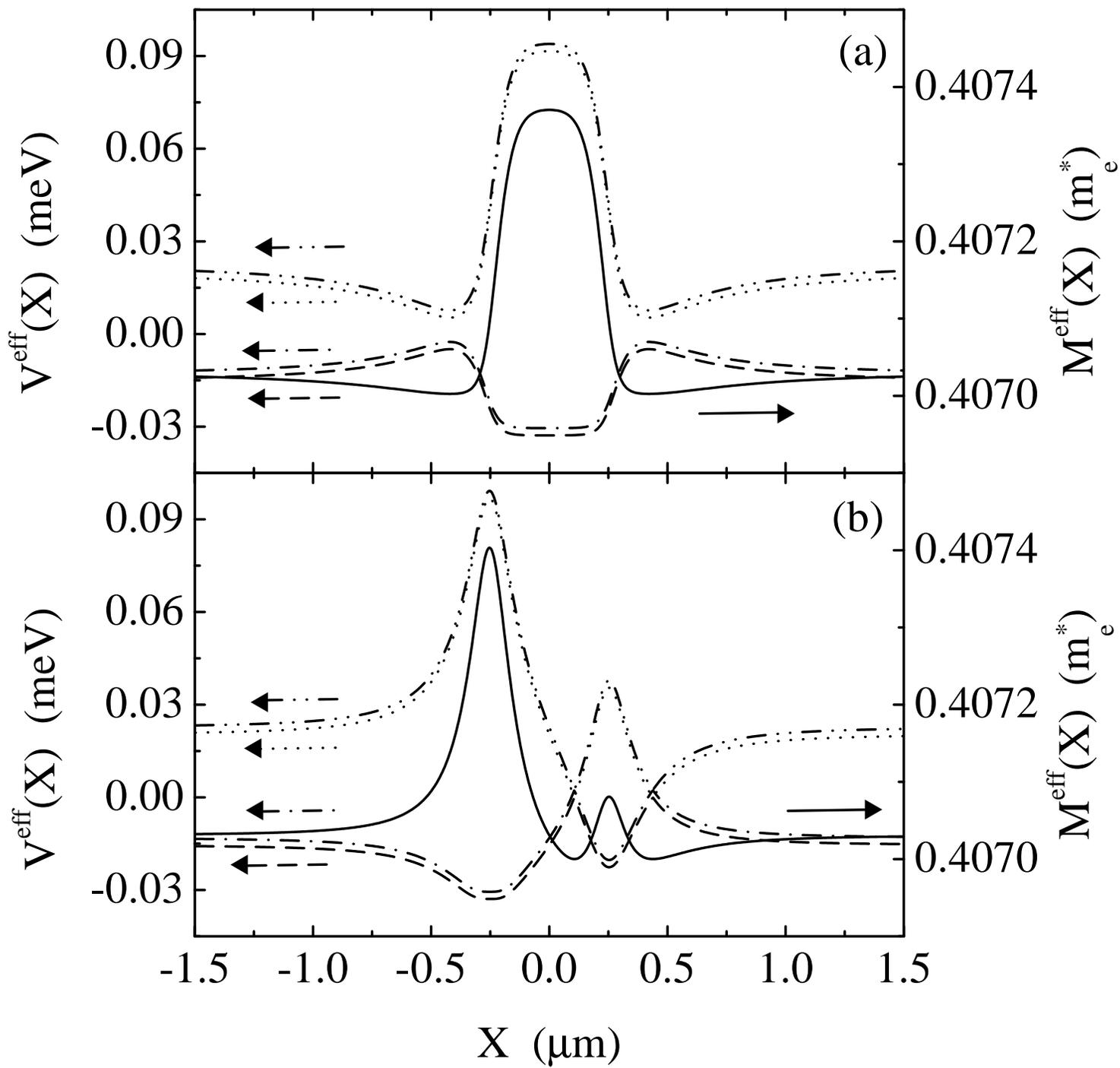
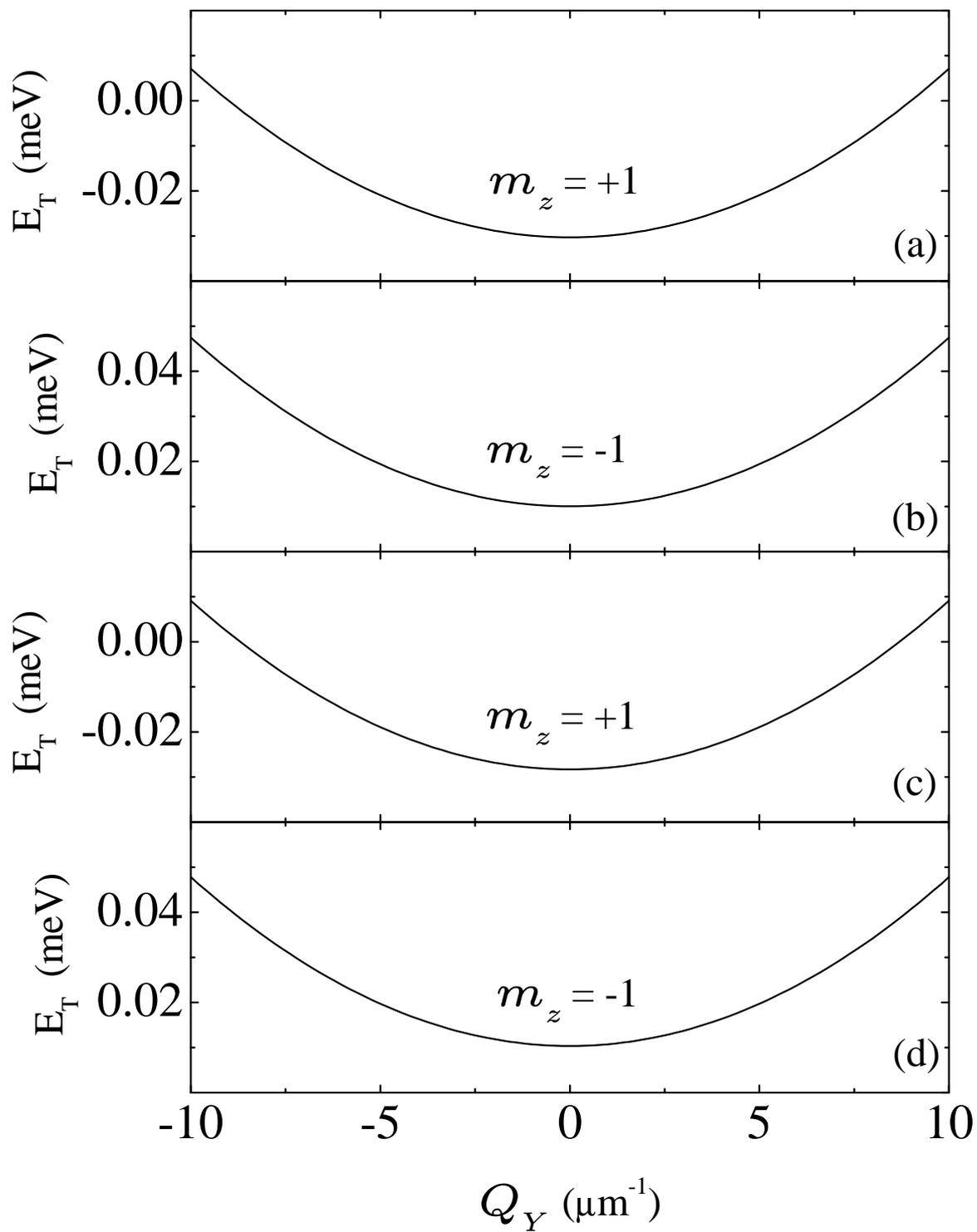
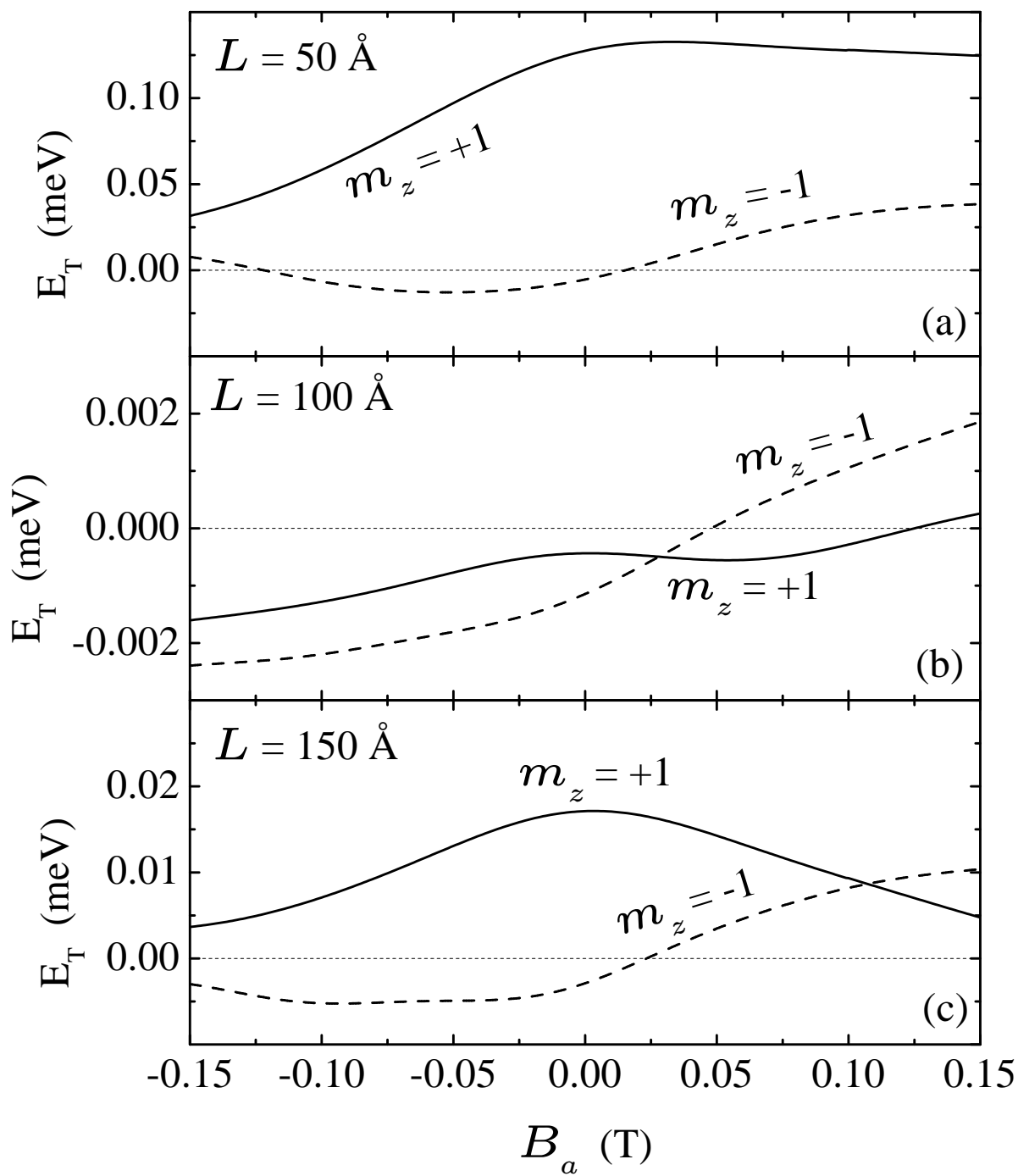
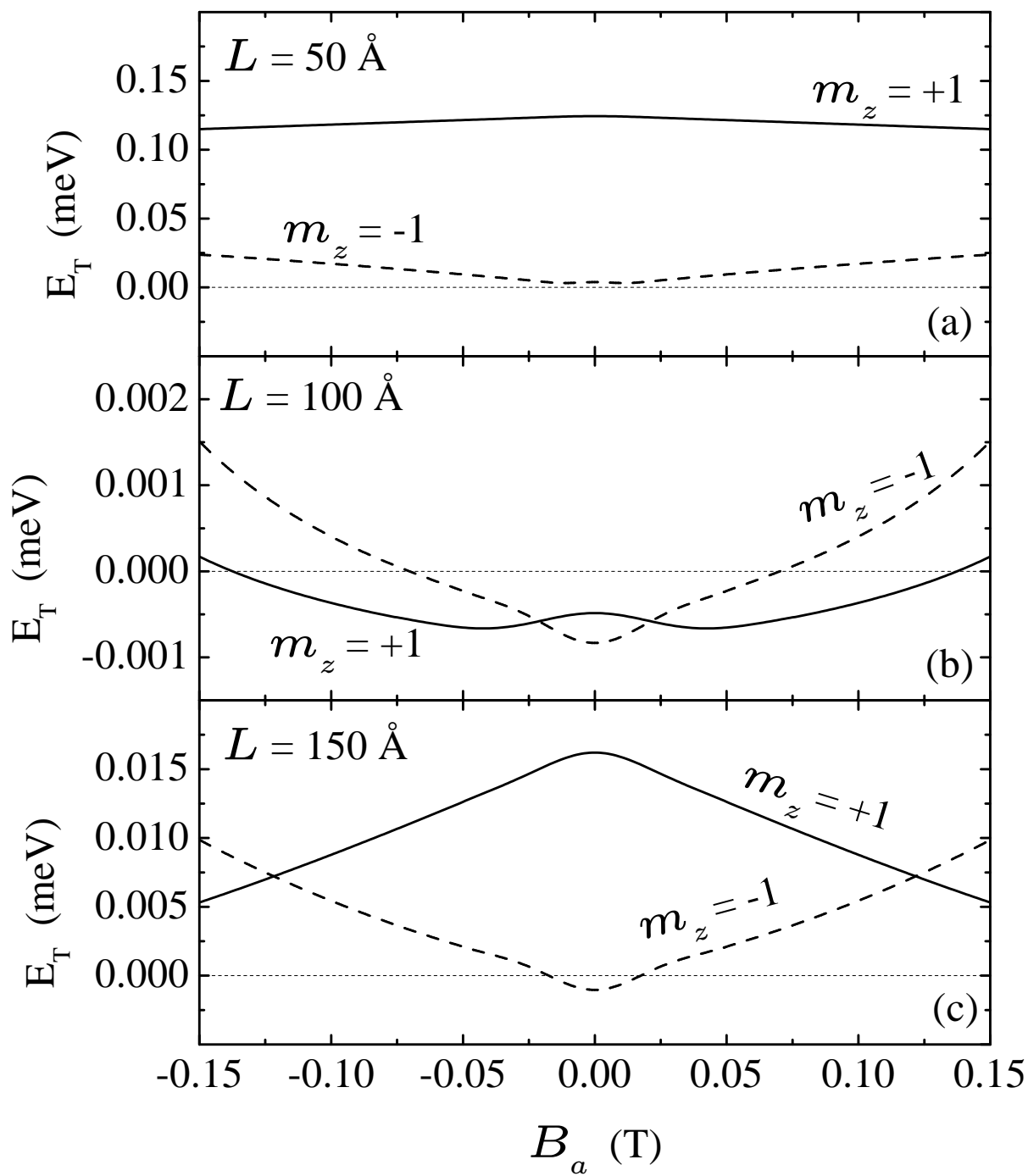
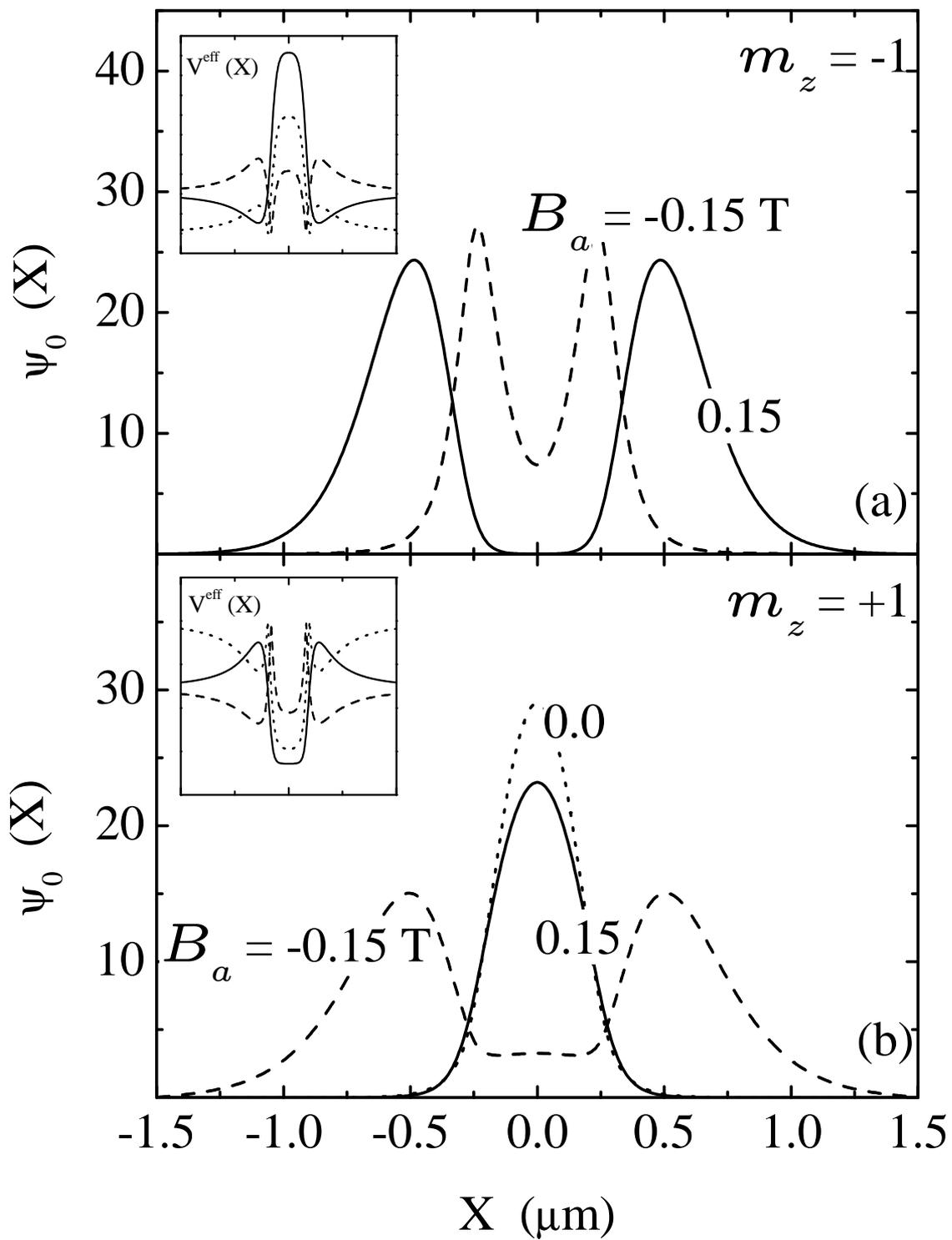
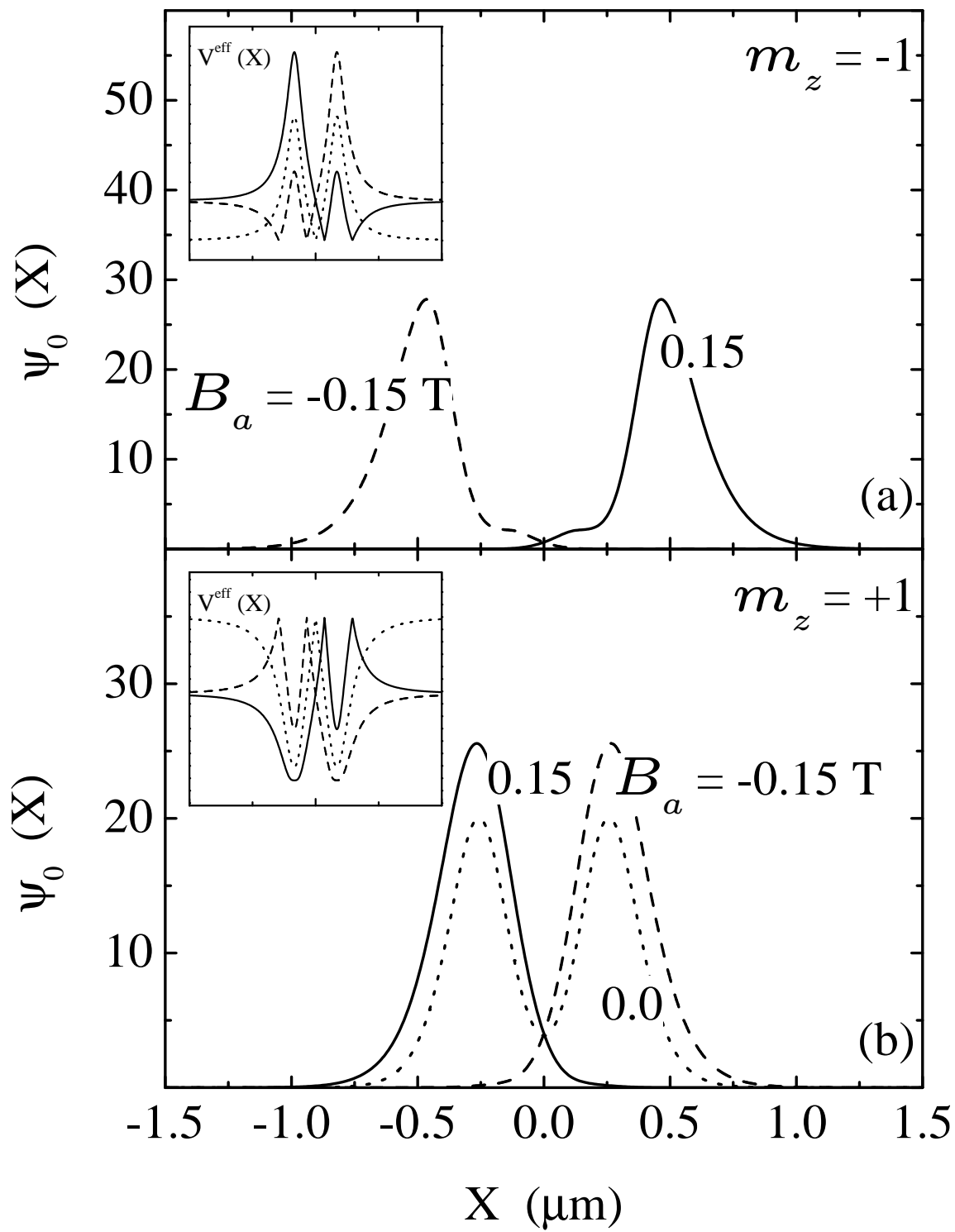
\end{document}